\documentclass[twocolumn]{aastex631}
\usepackage{float}

\newcommand\aastex{AAS\TeX}
\newcommand\latex{La\TeX}

\received{March 17, 2025}
\revised{June 19, 2025}

\begin{document}

\title{An Early Look at the Performance of IGRINS-2 at Gemini-North with Application to the ultrahot Jupiter, WASP-33~b 
}

\author[0009-0005-5145-5165]{Yeon-Ho Choi}
\affiliation{Korea Astronomy and Space Science Institute, 776 Daedeok-daero, Yuseong-gu, Daejeon 34055, Republic of Korea}
\affiliation{Department of Astronomy and Space Science, University of Science and Technology, 217 Gajeong-ro, Yuseong-gu, Daejeon 34113, Republic of Korea}

\author{Ueejeong Jeong}
\affiliation{Korea Astronomy and Space Science Institute, 776 Daedeok-daero, Yuseong-gu, Daejeon 34055, Republic of Korea}
\affiliation{University of Texas at Austin, 2515 Speedway, Stop C1400, Austin, Texas 78712-1205, USA}

\author[0000-0003-0894-7824]{Jae-Joon Lee}
\affiliation{Korea Astronomy and Space Science Institute, 776 Daedeok-daero, Yuseong-gu, Daejeon 34055, Republic of Korea}

\author[0000-0001-9263-3275]{Hyun-Jeong Kim}
\affiliation{Korea Astronomy and Space Science Institute, 776 Daedeok-daero, Yuseong-gu, Daejeon 34055, Republic of Korea}

\author[0000-0002-0418-5335]{Heeyoung Oh}
\affiliation{Korea Astronomy and Space Science Institute, 776 Daedeok-daero, Yuseong-gu, Daejeon 34055, Republic of Korea}

\author[0000-0001-9773-3080]{Chan Park}
\affiliation{Korea Astronomy and Space Science Institute, 776 Daedeok-daero, Yuseong-gu, Daejeon 34055, Republic of Korea}

\author[0000-0001-6876-8559]{Changwoo Kye}
\affiliation{Department of Physics and Astronomy, Seoul National University, Seoul 08826, Republic of Korea}

\author[0000-0002-1392-0768]{Luke Finnerty}
\affiliation{Department of Physics \& Astronomy, 430 Portola Plaza, University of California, Los Angeles, CA 90095, USA}


\author[0000-0002-2338-476X]{Micheal R. Line}
\affiliation{School of Earth and Space Exploration, Arizona State University, Tempe, AZ 85287, USA}

\author[0009-0005-4890-3326]{Krishna Kanumalla}
\affiliation{School of Earth and Space Exploration, Arizona State University, Tempe, AZ 85287, USA}

\author[0000-0002-9142-6378]{Jorge A. Sanchez}
\affiliation{School of Earth and Space Exploration, Arizona State University, Tempe, AZ 85287, USA}

\author[0000-0002-9946-5259]{Peter C. B. Smith}
\affiliation{School of Earth and Space Exploration, Arizona State University, Tempe, AZ 85287, USA}

\author{Sanghyuk Kim}
\affiliation{Korea Astronomy and Space Science Institute, 776 Daedeok-daero, Yuseong-gu, Daejeon 34055, Republic of Korea}

\author[0000-0003-0871-3665]{Hye-In Lee}
\affiliation{Korea Astronomy and Space Science Institute, 776 Daedeok-daero, Yuseong-gu, Daejeon 34055, Republic of Korea}

\author[0000-0001-8012-5871]{Woojin Park}
\affiliation{Korea Astronomy and Space Science Institute, 776 Daedeok-daero, Yuseong-gu, Daejeon 34055, Republic of Korea}

\author{Youngsam Yu}
\affiliation{Korea Astronomy and Space Science Institute, 776 Daedeok-daero, Yuseong-gu, Daejeon 34055, Republic of Korea}

\author{Yunjong Kim}
\affiliation{Korea Astronomy and Space Science Institute, 776 Daedeok-daero, Yuseong-gu, Daejeon 34055, Republic of Korea}

\author{Moo-Young Chun}
\affiliation{Korea Astronomy and Space Science Institute, 776 Daedeok-daero, Yuseong-gu, Daejeon 34055, Republic of Korea}

\author{Jae Sok Oh}
\affiliation{Korea Astronomy and Space Science Institute, 776 Daedeok-daero, Yuseong-gu, Daejeon 34055, Republic of Korea}

\author{Sungho Lee}
\affiliation{Korea Astronomy and Space Science Institute, 776 Daedeok-daero, Yuseong-gu, Daejeon 34055, Republic of Korea}

\author{Jeong-Gyun Jang}
\affiliation{Korea Astronomy and Space Science Institute, 776 Daedeok-daero, Yuseong-gu, Daejeon 34055, Republic of Korea}

\author{Bi-Ho Jang}
\affiliation{Korea Astronomy and Space Science Institute, 776 Daedeok-daero, Yuseong-gu, Daejeon 34055, Republic of Korea}

\author{Hyeon Cheol Seong}
\affiliation{Korea Astronomy and Space Science Institute, 776 Daedeok-daero, Yuseong-gu, Daejeon 34055, Republic of Korea}

\author{Cynthia B. Brooks}
\affiliation{University of Texas at Austin, 2515 Speedway, Stop C1400, Austin, Texas 78712-1205, USA}

\author[0000-0001-7875-6391]{Gregory N. Mace}
\affiliation{University of Texas at Austin, 2515 Speedway, Stop C1400, Austin, Texas 78712-1205, USA}

\author{Hanshin Lee}
\affiliation{University of Texas at Austin, 2515 Speedway, Stop C1400, Austin, Texas 78712-1205, USA}

\author{John M. Good}
\affiliation{University of Texas at Austin, 2515 Speedway, Stop C1400, Austin, Texas 78712-1205, USA}

\author[0000-0003-3577-3540]{Daniel T. Jaffe}
\affiliation{University of Texas at Austin, 2515 Speedway, Stop C1400, Austin, Texas 78712-1205, USA}

\author{Kang-Min Kim}
\affiliation{Korea Astronomy and Space Science Institute, 776 Daedeok-daero, Yuseong-gu, Daejeon 34055, Republic of Korea}

\author{In-Soo Yuk}
\affiliation{Korea Astronomy and Space Science Institute, 776 Daedeok-daero, Yuseong-gu, Daejeon 34055, Republic of Korea}

\author[0000-0002-2013-1273]{Narae Hwang}
\affiliation{Korea Astronomy and Space Science Institute, 776 Daedeok-daero, Yuseong-gu, Daejeon 34055, Republic of Korea}

\author[0000-0002-6982-7722]{Byeong-Gon Park}
\affiliation{Korea Astronomy and Space Science Institute, 776 Daedeok-daero, Yuseong-gu, Daejeon 34055, Republic of Korea}

\author[0000-0003-4770-688X]{Hwihyun Kim}
\affiliation{International Gemini Observatory/NSF NOIRLab, 950 N. Cherry Ave., Tucson, AZ 85719, USA}

\author{Brian Chinn}
\affiliation{International Gemini Observatory/NSF NOIRLab, Casilla 603, La Serena, Chile}

\author{Francisco Ramos}
\affiliation{International Gemini Observatory/NSF NOIRLab, Casilla 603, La Serena, Chile}

\author{Pablo Prado}
\affiliation{International Gemini Observatory/NSF NOIRLab, Casilla 603, La Serena, Chile}

\author{Ruben Diaz}
\affiliation{International Gemini Observatory/NSF NOIRLab, Casilla 603, La Serena, Chile}

\author{John White}
\affiliation{International Gemini Observatory/NSF NOIRLab, 670 N. A’ohoku Place, Hilo, Hawai’i, 96720, USA}

\author{Eduardo Tapia}
\affiliation{International Gemini Observatory/NSF NOIRLab, 670 N. A’ohoku Place, Hilo, Hawai’i, 96720, USA}

\author{Andres Olivares}
\affiliation{International Gemini Observatory/NSF NOIRLab, Casilla 603, La Serena, Chile}

\author{Valentina Oyarzun}
\affiliation{International Gemini Observatory/NSF NOIRLab, Casilla 603, La Serena, Chile}

\author{Emma Kurz}
\affiliation{International Gemini Observatory/NSF NOIRLab, 670 N. A’ohoku Place, Hilo, Hawai’i, 96720, USA}

\author{Hawi Stecher}
\affiliation{International Gemini Observatory/NSF NOIRLab, 670 N. A’ohoku Place, Hilo, Hawai’i, 96720, USA}

\author{Carlos Quiroz}
\affiliation{International Gemini Observatory/NSF NOIRLab, Casilla 603, La Serena, Chile}

\author{Ignacio Arriagada}
\affiliation{International Gemini Observatory/NSF NOIRLab, Casilla 603, La Serena, Chile}

\author{Thomas L. Hayward}
\affiliation{International Gemini Observatory/NSF NOIRLab, Casilla 603, La Serena, Chile}

\author[0000-0002-2536-1633]{Hyewon Suh}
\affiliation{International Gemini Observatory/NSF NOIRLab, 670 N. A’ohoku Place, Hilo, Hawai’i, 96720, USA}

\author{Jen Miller}
\affiliation{International Gemini Observatory/NSF NOIRLab, 670 N. A’ohoku Place, Hilo, Hawai’i, 96720, USA}

\author{Siyi Xu}
\affiliation{International Gemini Observatory/NSF NOIRLab, 670 N. A’ohoku Place, Hilo, Hawai’i, 96720, USA}

\author[0000-0002-6822-2254]{Emanuele Paolo Farina}
\affiliation{International Gemini Observatory/NSF NOIRLab, 670 N. A’ohoku Place, Hilo, Hawai’i, 96720, USA}

\author{Charlie Figura}
\affiliation{International Gemini Observatory/NSF NOIRLab, 670 N. A’ohoku Place, Hilo, Hawai’i, 96720, USA}

\author[0000-0003-4603-556X]{Teo Mocnik}
\affiliation{International Gemini Observatory/NSF NOIRLab, 670 N. A’ohoku Place, Hilo, Hawai’i, 96720, USA}

\author[0000-0003-4236-6927]{Zachary Hartman}
\affiliation{International Gemini Observatory/NSF NOIRLab, 670 N. A’ohoku Place, Hilo, Hawai’i, 96720, USA}
\affiliation{NASA Ames Research Center, Moffett Field, CA 94035, USA}

\author[0000-0002-6529-202X]{Mark Rawlings}
\affiliation{International Gemini Observatory/NSF NOIRLab, 670 N. A’ohoku Place, Hilo, Hawai’i, 96720, USA}

\author{Andrew Stephens}
\affiliation{International Gemini Observatory/NSF NOIRLab, 670 N. A’ohoku Place, Hilo, Hawai’i, 96720, USA}

\author[0000-0002-5665-376X]{Bryan Miller}
\affiliation{International Gemini Observatory/NSF NOIRLab, Casilla 603, La Serena, Chile}

\author[0000-0002-6633-7891]{Kathleen Labrie}
\affiliation{International Gemini Observatory/NSF NOIRLab, 670 N. A’ohoku Place, Hilo, Hawai’i, 96720, USA}

\author{Paul Hirst}
\affiliation{International Gemini Observatory/NSF NOIRLab, 670 N. A’ohoku Place, Hilo, Hawai’i, 96720, USA}

\author{Byeong-Cheol Lee}
\affiliation{Korea Astronomy and Space Science Institute, 776 Daedeok-daero, Yuseong-gu, Daejeon 34055, Republic of Korea}
\affiliation{Department of Astronomy and Space Science, University of Science and Technology, 217 Gajeong-ro, Yuseong-gu, Daejeon 34113, Republic of Korea}

\correspondingauthor{Byeong-Cheol Lee}
\email{bclee@kasi.re.kr}



\begin{abstract}

Ground-based high-resolution spectroscopy enables precise molecular detections and velocity-resolved atmospheric dynamics, offering a distinct advantage over low-resolution methods for exoplanetary atmospheric studies.
IGRINS-2, the successor to IGRINS, features improved throughput and enhanced sensitivity to carbon monoxide by shifting its \textit{K}-band coverage by 36 nm to longer wavelengths. IGRINS is a near-infrared high-resolution spectrograph mounted at McDonald, Lowell, and Gemini-South observatories. Our order-drop test shows this added range improves the CO cross-correlation signal-to-noise ratio (SNR) by $\sim$2–-3$\%$, confirming a measurable but modest sensitivity gain. To evaluate its performance, we attempt to investigate the atmospheric characteristics of WASP-33 b. Observations were conducted on 2024 January 7 for a total of 2.43 hours; This includes 1.46 hours in the pre-eclipse phase to capture the planet’s thermal emission spectrum. 
We successfully detect clear cross-correlation signals from molecular species in the dayside atmosphere of WASP-33 b with a combined SNR of 7.4. More specifically, we capture CO, H$_{2}$O, and OH with SNRs of 6.3, 4.7, and 4.2, respectively. These results are consistent with previous studies and demonstrate that IGRINS-2 is well-suited for detailed investigation of exoplanetary atmospheres.
We anticipate that future observations with IGRINS-2 will further advance our understanding of exoplanetary atmospheres.



\end{abstract}



\section{Introduction} \label{sec:intro}


High-resolution infrared spectroscopy (R $= \lambda/\Delta \lambda >$ 15,000) has become a cornerstone for characterizing exoplanetary atmospheres, offering unique advantages over low-resolution spectroscopy. 
Unlike low-resolution spectroscopy, which primarily probes higher layers of an atmosphere, high-resolution spectroscopy enables the detection of molecular species across a broader range of pressures, providing insights into thermal profiles and atmospheric dynamics \citep{Snellen2010}.
Additionally, high-resolution spectroscopy is highly sensitive to velocity-resolved information, allowing for the study of wind patterns and rotational dynamics \citep{Wardenier2024}. These capabilities offer a deeper understanding of atmospheric chemistry and planetary formation histories \citep{Oberg2011, Booth2017, Chachan2023}. High-resolution cross-correlation spectroscopy (HRCCS), in particular, has become an essential tool for exploring exoplanetary atmospheres. By aligning observed spectra with theoretical templates, HRCCS allows for the detection of molecular and atomic species and provides insights into thermal and velocity structures. This technique has been successfully applied to various orbital geometries (i.e., transmission and thermal emission), enabling a comprehensive investigation of exoplanetary atmospheres \citep{Snellen2010, deKok2013, Brogi2023}.

The Immersion GRating INfrared Spectrograph 2 \citep[IGRINS-2;][]{Lee2022, Oh2024} has been recently integrated as a facility instrument at the 8.1 m Gemini-North Observatory. IGRINS-2 builds upon the success of its predecessor IGRINS \citep{Yuk2010, Park2014, Mace2016, Mace2018} by offering high-resolution spectroscopy in the near-infrared \textit{H} (1.442--1.836~\micron, R~$\sim$~50,000) and \textit{K} (1.892--2.554~\micron, R~$\sim$~45,000) bands.
With these specifications, the instrument has a particular advantage for HRCCS in probing the chemical composition of hot Jupiters due to its wide simultaneous wavelength coverage, as demonstrated by previous high-resolution studies using IGRINS \citep[e.g.,][]{Line2021}, Gemini/MAROON-X \citep[e.g.,][]{Pelletier2023}, VLT/ESPRESSO \citep[e.g.,][]{Borsa2021}, VLT/CRIRES+ \citep[e.g.,][]{Yan2023}, and TNG/GIANO \citep[e.g.,][]{Giacobbe2021}. Additionally, with IGRINS-2 now operating in the northern hemisphere, we not only gain access to a new population of northern targets, which have not been explored by IGRINS at Gemini-South\footnote{IGRINS was operated at Gemini-South as a visiting instrument until April 2024 and is now in operation at McDonald Observatory}, but also open possibilities for complementary follow-up observations using other high-resolution spectrographs such as Gemini-North/MAROON-X~\citep{MAROONX}, LBT/PEPSI~\citep{PEPSI}, Subaru/IRD~\citep{IRD}, Subaru/HDS~\citep{HDS}, and Keck I/HIRES~\citep{HIRES}, Keck II/NIRSPEC~\citep{NIRSPEC, NIRSPEC2}.

\begin{figure*}[ht!] \centering
\includegraphics[width=0.99\textwidth]{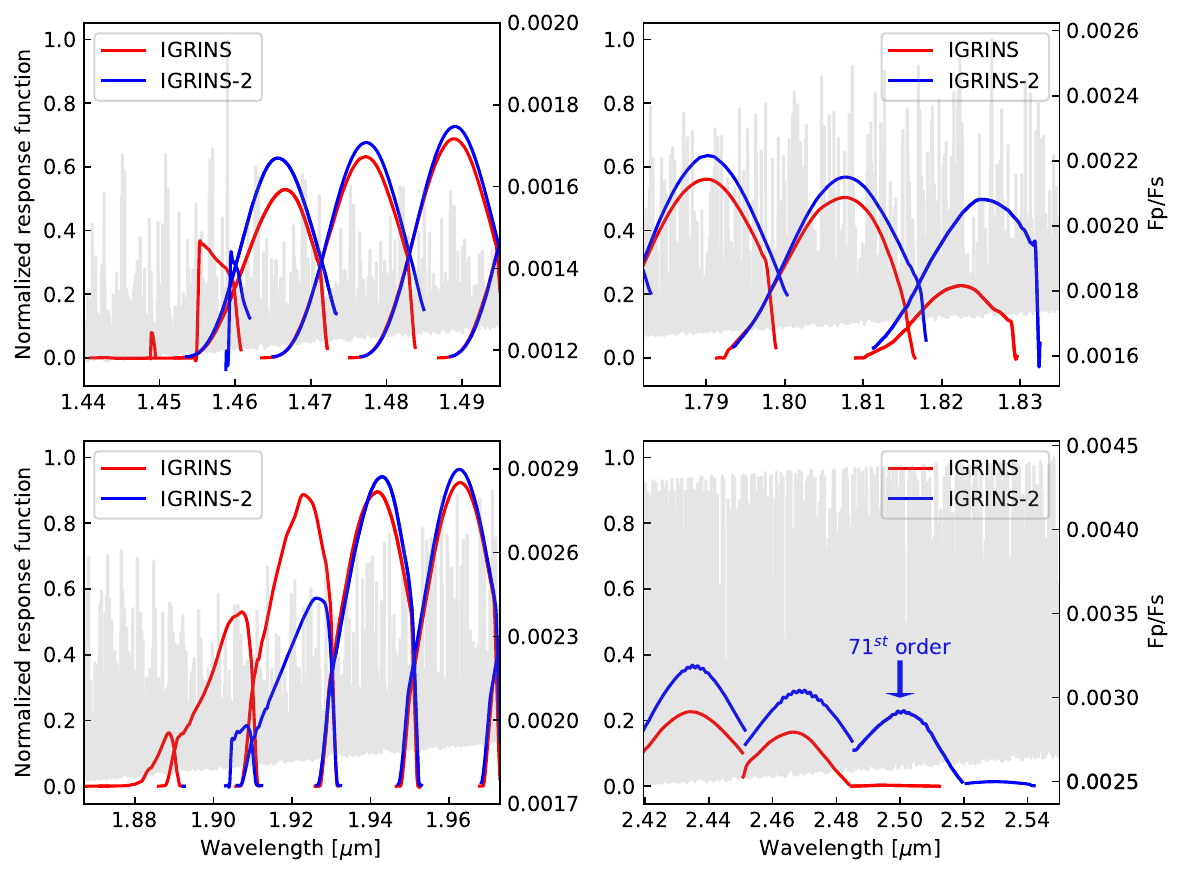}

\caption{
Comparison of normalized response functions for IGRINS (red) and IGRINS-2 (blue) obtained from flat lamp frames in the \textit{H} \& \textit{K} bands with the atmospheric model (gray) used for the analysis. Spectra are obtained from calibration systems after correcting the minor changes in optics. Top panels: the normalized response function of each spectrograph in the \textit{H} band. IGRINS-2 exhibits improved sensitivity over IGRINS in the entire wavelength range. Bottom panels: the normalized response function of each spectrograph in the \textit{K} band. These panels show the significant shift in the \textit{K} band to longer wavelengths in IGRINS-2, covering more CO lines and improving throughput at longer wavelengths.}
\label{fig:inst}
\end{figure*}

WASP-33~b, an ultrahot Jupiter with an equilibrium temperature of approximately 2781 K \citep{Chakrabarty2019}, serves as an ideal target for testing IGRINS-2’s capabilities in HRCCS due to its brightness and high equilibrium temperature. Planets in this temperature regime have been investigated with high-resolution spectroscopy in both the infrared and optical wavelengths, revealing a variety of atomic and molecular species such as CO, H$_{2}$O, OH, and Fe \citep[e.g.,][]{Yan2020, Cabot2021, Brogi2023, Ramkumar2023, Damasceno2024}. The WASP-33 system is a well-studied system with key properties including a short orbital period of 1.22 days \citep{Chakrabarty2019} and orbiting a bright A5 star ($V=8.14~\rm{mag}$, $K=7.46~\rm{mag}$).
Discovered by \citet{Christian2006} through SuperWASP transit photometry, WASP-33~b orbits its host star in a near-polar retrograde orbit, likely influenced by dynamical interactions such as the Kozai-Lidov effect \citep{Naoz2011}. 

Previous high-resolution spectroscopy studies have revealed a thermally inverted atmosphere on the dayside of WASP-33~b, characterized by significant photodissociation of H$_{2}$O and the detection of various molecules and atomic species. Emission spectroscopy has identified CO, OH, TiO, \ion{Fe}{1}, \ion{Si}{1}, \ion{Ti}{1}, \ion{V}{1} as prominent components in the dayside atmosphere \citep{Nugroho2017, Nugroho2020, Nugroho2021, Cont2021, Cont2022a, Cont2022b, Yan2022, Sluijs2023}, while transmission spectroscopy has detected hydrogen Balmer lines and \ion{Ca}{2} in the terminator region \citep{Yan2019, Yan2021}. These findings suggest a complex atmospheric structure, with early inferences pointing to a thermal inversion on the dayside. \citet{Yang2024} extend this understanding by suggesting that thermal inversion also occurs in the terminator region, offering new insights into the atmospheric dynamics and chemical processes of WASP-33~b.

The recent study by \citet[][hereafter, F23]{Finnerty2023} using the Keck Planet Imager and Characterizer (KPIC; R $\sim$ 35,000) provided comprehensive insights into the atmospheric composition of WASP-33~b. 
Their analysis validated a thermally inverted dayside atmosphere with significant detections of CO, H$_{2}$O, and OH in emissions.
Following the detection of molecular species, Bayesian atmospheric retrievals \citep{BrogiLine2019} were performed to place quantitative constraints on the composition of WASP-33~b’s dayside atmosphere. The model fits indicated a carbon- and metal-enriched environment, with a gas-phase C/O ratio of 0.8$\pm$0.2 and evidence for a metal-enriched atmosphere (2--15 $\times$ solar). 
These findings are consistent with formation scenarios that involve accretion near the CO or CO$_{2}$ snow lines \citep{Oberg2011} or enrichment by carbon-rich grains interior to the H$_{2}$O snowline \citep{Chachan2023}, as suggested by F23. 
Additionally, F23 found tentative evidence for a $^{12}$CO/$^{13}$CO ratio of $\sim$ 50 consistent with values expected in protoplanetary disks.

In this paper, we present an early look at the performance of IGRINS-2 for the atmospheric characterization of an exoplanet using WASP-33~b. Our findings demonstrate that IGRINS-2 is capable of detecting molecular signatures, showcasing its potential for future exoplanetary atmospheric studies. We provide a detailed account of our observational methods, data reduction techniques, and analysis, illustrating the instrument's suitability for high-resolution spectroscopic investigations.

\section{Upgrades on IGRINS-2}

The upgrades from IGRINS to IGRINS-2 include enhanced sensitivity and a shift in spectral coverage, which are critical for HRCCS of exoplanetary atmospheres. This shift resulted from the process of aligning the echellogram on the detector by adjusting the main disperser, the immersion grating. 
The shift toward longer wavelengths was not deliberate but arose naturally during the optical alignment and optimization process of the spectrograph \citep{Oh2024}. Specifically, physical adjustments were made to the spectrograph’s main disperser, the immersion grating, and the detector to optimize optical performance within the required \textit{K}-band range of 1.96--2.46 \micron. As part of this process, additional adjustments were implemented to minimize the impact of detector artifacts, such as bad pixels, on the spectra. This technical optimization resulted in a slightly shifted coverage at longer wavelengths in \textit{K}-band.
To isolate instrumental differences from site-specific effects (e.g., telescope throughput, atmospheric transmission), we compared flat lamp frames from the calibration systems of both instruments. 
The major changes in optics include replacing the mirror of the IGRINS’s calibration system with a beam splitter in IGRINS-2 and using a dichroic mirror with enhanced response at the edges of each channel in the spectrograph.
We corrected these differences by interpolating and normalizing the response function based on the transmittance of each optical component, though these corrections had an insignificant impact on the final spectral response (see Figure~\ref{fig:inst}).

\subsection{\textit{K}-Band Spectral Shift}
One of the most significant changes in IGRINS-2 is the shifts of the H and \textit{K}-band spectra to longer wavelengths. Specifically, the \textit{K} band now extends approximately 36 nm further to longer wavelengths compared to IGRINS. This shift could be crucial because it captures additional spectral features, particularly CO molecular lines that begin to appear around 2.3~\micron. By covering more CO lines in the \textit{K} band, as shown in the bottom right panel of Figure 1, IGRINS-2 improves the detection and characterization of molecular features in exoplanetary atmospheres, resulting in an increased signal in the cross-correlation analysis. The impact of including this extended spectral range on CO, particularly the 71$^{\rm st}$ order (2.48--2.52 µm), which is near the longest spectral order, is tested and discussed further in Section 4.2.

\subsection{Sensitivity Enhancements}
Another significant advancement in IGRINS-2 is the improvement in the quantum efficiency (QE) of the detector in the \textit{K} band. The QE of IGRINS in the \textit{K} band was approximately 80--85\% \citep{Yuk2010}, whereas IGRINS-2 achieved an average QE of 96\% (C. Park et al. in preparation). This increase in QE results in more efficient photon collection, particularly at longer wavelengths, enhancing the overall sensitivity of the instrument and improving the detection of weaker molecular features.

The impact of this enhanced QE is evident in the noticeable increase in response functions obtained in the extended portion of the \textit{K} band (see the bottom right panel of Figure.~\ref{fig:inst}). This improvement in sensitivity is crucial for detecting faint molecular signals in exoplanet atmospheres, as the planet's signal-to-noise ratio is directly proportional to the stellar signal-to-noise ratio (SNR), as described by \citet{Birkby2018}. Together, these advancements, particularly in the longer wavelength region, should enable more robust atmospheric detections, especially for species such as CO, which are prominent in the \textit{K} band.

\begin{figure}[ht!] \centering
\includegraphics[width=0.47\textwidth]{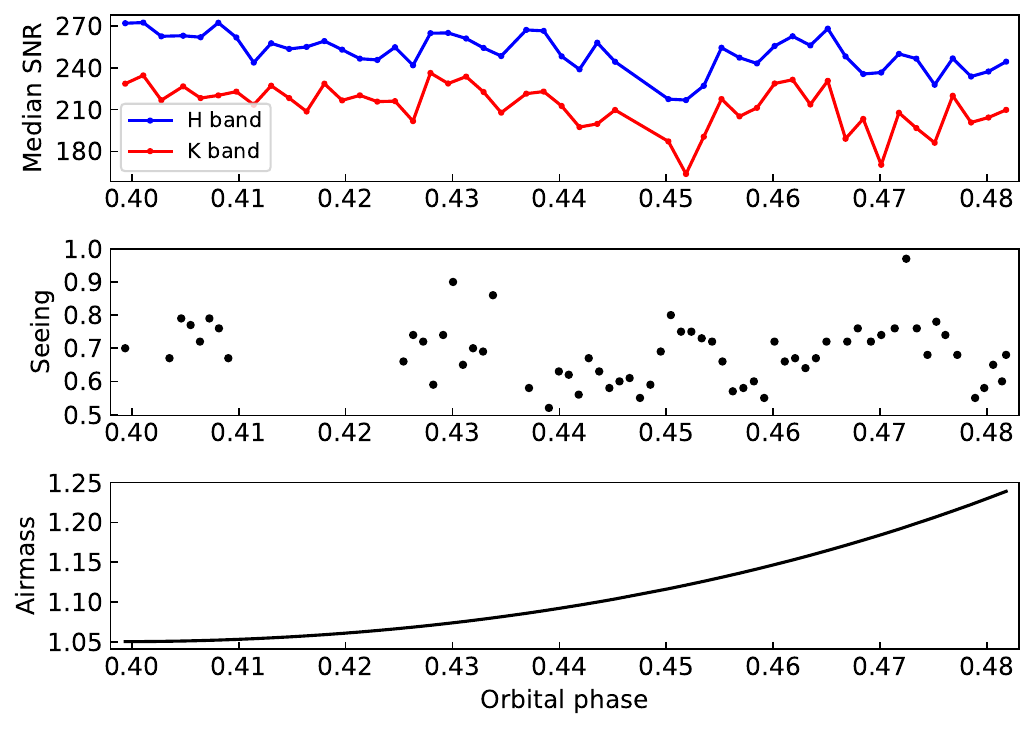}
\caption{Top panel: the median SNR of each band. SNRs are derived from an exposure time of 80 seconds while observing an A5 star ($K=7.46~\rm{mag}$). Middle panel: seeing variations obtained from DIMM. Bottom panel: airmass variations during WASP-33~b observations.}
\label{fig:obs}
\end{figure}

\begin{figure}[!t] \centering
\includegraphics[width=0.47\textwidth]{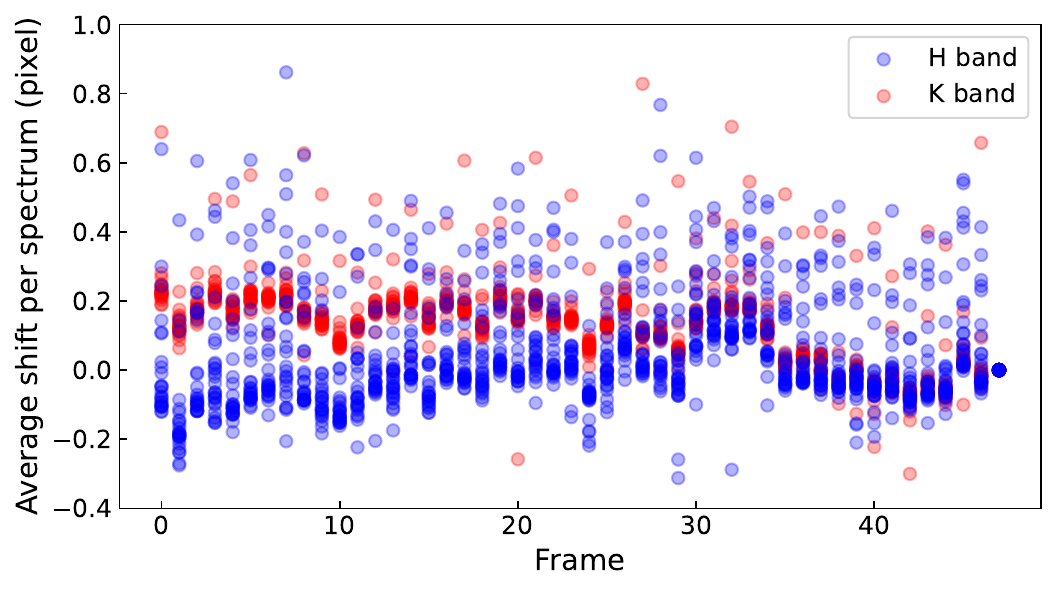}

\caption{Results of the secondary wavelength calibration. Each point represents the wavelength shift for each spectral order. Since the night-sky frame used for wavelength calibration was taken after our observations, the last frame serves as the reference.
}
\label{fig:wavecal}
\end{figure}

\section{Observations and Data reduction} \label{sec:obs}
\subsection{Observations}
On UT date 2024 January 7, we observed WASP-33~b as part of the test observations of the IGRINS-2 attached to the Gemini-North telescope. The observations of WASP-33~b were designed to probe the dayside thermal emission spectrum and benchmark its atmospheric characterization capabilities, evaluating the effectiveness of IGRINS-2 for high-resolution exoplanet spectroscopy.
We obtained a continuous sequence of 48 AB pairs over 2.43 hours covering the pre-eclipse orbital phases (0.399 $< \phi <$ 0.481) with the secondary eclipse of WASP-33~b beginning at orbital phase 0.451. 
The integrated time for a single AB pair was 80 seconds. 

Figure~\ref{fig:obs} shows the SNR, seeing, and airmass variations during the observations. Throughout the observation period, the relative humidity remained stable at approximately 9\%, as recorded by the Maunakea Weather Center\footnote[2]{\url{http://mkwc.ifa.hawaii.edu/archive/}} at the Canada-France-Hawaii Telescope station, which is located near Gemini-North. Seeing, obtained from the Differential Image Motion Monitor (DIMM), varied between $0\farcs52$ and $0\farcs97$, with an average seeing of approximately $0\farcs68$. 

\subsection{Data reduction}

The data reduction was performed using the IGRINS Pipeline Package Version 3.0 \citep[PLP;][]{Kaplan2024}, which was initially created for the original IGRINS. The \texttt{igrins2-dev} branch\footnote[3]{\url{https://github.com/igrins/plp/tree/igrins2-dev}} of the pipeline, which contains the necessary modifications for IGRINS-2, was utilized. Flat-fielding was applied to the raw data to adjust for pixel-to-pixel differences. Bad pixels were identified and removed from both flat-off and flat-on images. One-dimensional spectra were extracted using a modified version of \citet{Horne1986} optimal extraction techniques. The final result consists of one-dimensional spectra for each diffraction order.

After initial data reduction, we used the \texttt{IGRINS\_transit}\footnote[4]{\url{https://github.com/meganmansfield/IGRINS_transit}} \citep{Mansfield2024a} package for secondary wavelength calibration and barycentric-velocity correction which was described in \citet{Line2021}.
\texttt{IGRINS$\_$transit} also includes routines for the data cube production, pixel trimming, and exclusion of bad orders. We trimmed 100 pixels from the edges of each order due to their low SNRs and discarded orders at the edges of each band that had low transmittance and strong telluric contamination during calibration. As a result, five orders were discarded out of the total 53 orders: the longest order (2.520--2.554~\micron) and the shortest three orders (1.866--1.932~\micron) in the \textit{K} band, as well as the shortest order (1.441--1.461~\micron) in the \textit{H} band. Each remaining order contains 1848 pixels. Then, a secondary wavelength calibration using \texttt{scipy.curve\_fit} was applied to refine the wavelength solution. This corrected any residual shifts in the wavelength calibration across the spectral orders. 
Figure~\ref{fig:wavecal} shows the resulting wavelength calibration shifts over time for both the \textit{H} \& \textit{K} bands, where each point represents the shift for an individual spectral order.

To minimize noise contributions from stellar variability and instrumental systematics, we excluded data obtained after the start of the secondary eclipse from subsequent analyses. Specifically, spectra taken during and after the eclipse ingress were removed. While PCA-based methods like SVD typically benefit from larger datasets, including spectra with no expected planetary signal (e.g., those during eclipse) can introduce non-planetary variance into the analysis, thereby suppressing the planetary signal in the decomposition. Excluding these data thus effectively reduces irrelevant variability, enhancing detection significance. The complete dataset (including eclipse observations) was preserved for trail plot visualization to maintain full orbital phase context.

The data cube, produced by stacking the time series spectra for each order and pixel, contains a mix of the stellar spectrum, telluric absorption, planet signal, and photon noise. To extract the planet signal, it is crucial to isolate and remove the stationary stellar and telluric lines, which remain quasi-stationary during observations, while the planet’s spectrum shifts due to its orbital motion. This can be achieved through singular value decomposition (SVD), which decomposes the time-series data into principal components \citep[see][]{Birkby2013, deKok2013, Giacobbe2021, Line2021}. The dominant stationary features are removed, leaving the planet signal in the residuals. The optimal number of SVD components to remove can vary depending on observing conditions and the level of telluric absorption. 
Following the approach described by F23, we performed cross-correlation analyses using different numbers of singular values (SVs). We found that using four SVs resulted in the highest SNR. However, the derived signals were robust to small changes in SV selection; specifically, the resulting SNR varied minimally (within one sigma) when using three to five SVs, confirming the stability of our detection.

\begin{figure*}[!ht] \centering
\includegraphics[width=0.96\textwidth]{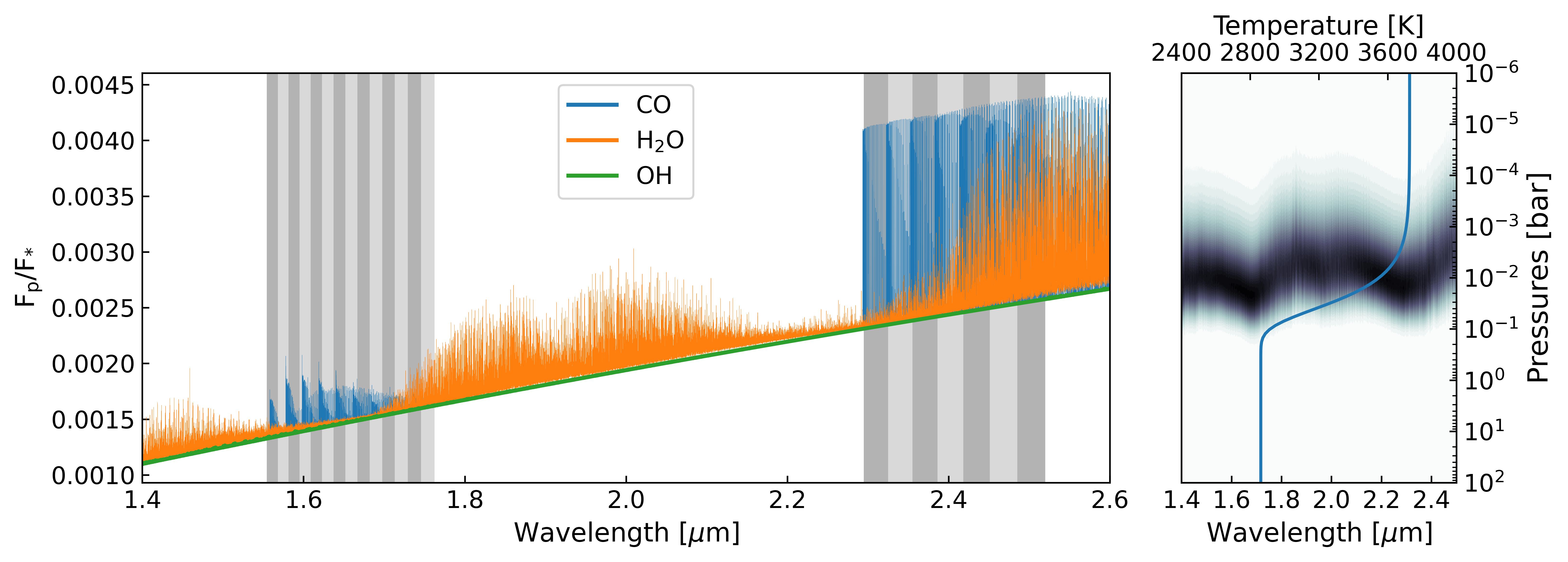}

\caption{
Atmospheric model spectra for WASP-33~b’s dayside, showing contributions from individual molecular species (CO, H$_{2}$O, and OH). The grey-shaded boxes represent the spectral orders used for CO detection. The right panel shows the corresponding P-T profile and emission contribution functions. The models were generated with petitRADTRANS using a Guillot P-T profile.
}
\label{fig:models}
\end{figure*}

\section{Molecular Signal Detection via Cross-correlation}

\subsection{Atmospheric models}


To perform cross-correlation analysis, we generated thermal emission spectra for WASP-33~b with a resolving power of 250,000 using the \texttt{petitRADTRANS} package \citep{Molliere2019, Molliere2020}. 
The atmospheric structure is modeled across a pressure range from 10$^{-6}$ to 10$^{2}$ bar, divided into 100 layers with uniform spacing in logarithmic pressure. Models for each molecule and pressure-temperature (P-T) profile used in the analysis are shown in Figure~\ref{fig:models}. Consistent with several previous studies, we assumed a thermal inversion profile for the dayside atmosphere \citep{Nugroho2021, van_Sluijs2023, Yan2022, Finnerty2023} and we adopted Guillot P-T profile \citep{Guillot2010}. Molecular species included in the models were H$_{2}$O \citep{Rothman2010}, CO \citep{Li2015}, and OH \citep{Gordon2022} from HITRAN/HITEMP, chosen for following the results from F23.
These models were built based on key parameters from F23, including reference pressure, mass mixing ratios (MMRs), and the Guillot P-T profile parameters $\gamma$, and $\kappa$ IR \citep[][private communication]{Finnerty2023}. Stellar and planetary properties were adopted from \citet{Chakrabarty2019}, and a PHOENIX model was used for the stellar template \citep{Husser2013}. The stellar template is then smoothed to approximate the stellar continuum. The parameters used for the atmospheric models are summarized in Table~\ref{atmos}.

\begin{table}
\caption{Parameters used for atmospheric model}
\label{atmos}
\begin{tabular}{lc}
\hline\hline
Parameter & Value \\
\hline
Stellar effective temperature & 7430 \rm{K} $^{a}$ \\ 
Stellar mass   & 1.495 ${M}_{\rm Sun}$ $^{a}$\\
Stellar radius & 1.444 ${R}_{\rm Sun}$ $^{a}$\\
Planetary mass & 2.093 ${M}_{\rm Jup}$ $^{a}$ \\
Planetary radius & 1.593 ${R}_{\rm Jup}$ $^{a}$ \\
Orbital period & 1.21987 days $^{a}$ \\
Transit duration & 2.854 hours $^{a}$ \\
{\rm K}${_{p}}$ & $\rm 226~km~s^{-1}$ $^{b}$ \\
Top-of-atmosphere temperature & 3200~\rm{K} \\
Intrinsic temperature & 50~\rm{K} \\ 
log MMR$_{\rm CO}$ & $-2.4$ $^{c}$ \\
log MMR$_{\rm H_{2}O}$ & $-4.4$ $^{c}$ \\
log MMR$_{\rm OH}$ & $-6.8$ $^{c}$ \\
log $\gamma$ & 0.5 $^{c}$ \\ 
log $\kappa$ IR & $-2.6$ $^{c}$ \\
CO ratio & 0.8 $^{c}$ \\
log reference pressure & $-2.8$ bar \\
\hline
\end{tabular}
\raggedright \\
$^{a}$ \citet{Chakrabarty2019} \\
$^{b}$ \citet{Finnerty2023} \\
$^{c}$ private communication

\end{table}

\begin{figure}[!th] \centering
\includegraphics[width=0.48\textwidth]{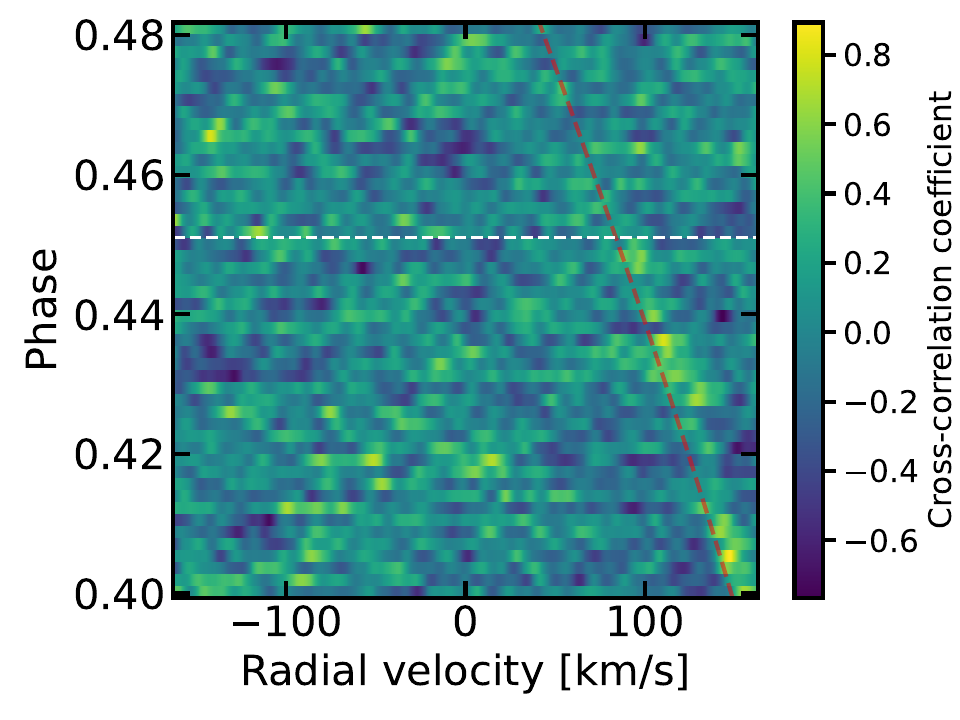}
\includegraphics[width=0.5\textwidth]{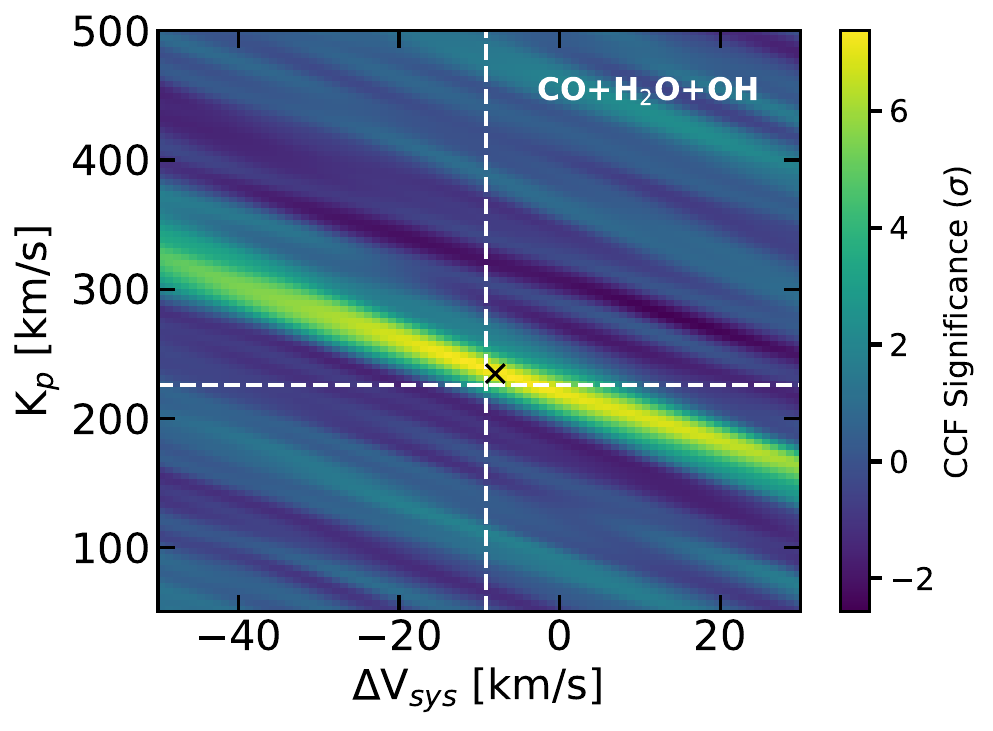}

\caption{
Detection of molecules (CO+H$_{2}$O+OH) in WASP-33~b's dayside atmosphere. Top panel: trail plot showing cross-correlation coefficients calculated from the post-SVD residual data using the composite atmospheric model of WASP-33~b. The dashed red line indicates the expected radial velocity of the planet ({\rm K}$_{p} = 226.0 \ {\rm km~s^{-1}}$), while the white horizontal dashed line indicates the ingress phase ($\phi = 0.451$) of WASP-33~b. Bottom panel: ${\rm K}_{p}$-$\Delta {\rm V}_{sys}$ map of SNR detection with a peak value of 7.4. The peak position (black cross) is well consistent with the expected planetary radial velocity and system velocity ({\rm V}${_{sys} = -9.2 \ {\rm km~s^{-1}}}$ from \citet{Gontcharov2006}), indicated by the dashed white lines.}
\label{fig:CCF_all}
\end{figure}

\bigskip
\bigskip

\subsection{Cross-correlation maps}

To validate the performance of IGRINS-2 with HRCCS, we cross-correlated the post-SVD residual spectra with atmospheric models using the Pearson correlation coefficient. These normalized correlation coefficients measure how well the variations in the observed spectra match those of the shifted model spectra. The atmospheric models were convolved with the IGRINS-2 instrumental profile to match the resolving power of the observations ($\rm{R}=45,000$). The cross-correlation function (CCF) was computed by systematically shifting the model spectra across a range of radial velocities and calculating the correlation at each step. We followed the same approach described in \citet{BrogiLine2019} and \citet{Line2021} that is used to detect the atmospheric signal using IGRINS \citep[see][]{Brogi2023, Kanumalla2024, Smith2024a, Mansfield2024b}.
Frames taken during the secondary eclipse, when the planet is hidden behind its host star, were not included when generating  ${\rm K}_{p}$-${\rm V}_{sys}$ maps.

\begin{figure*}[htbp] \centering
\includegraphics[width=1.0\textwidth]{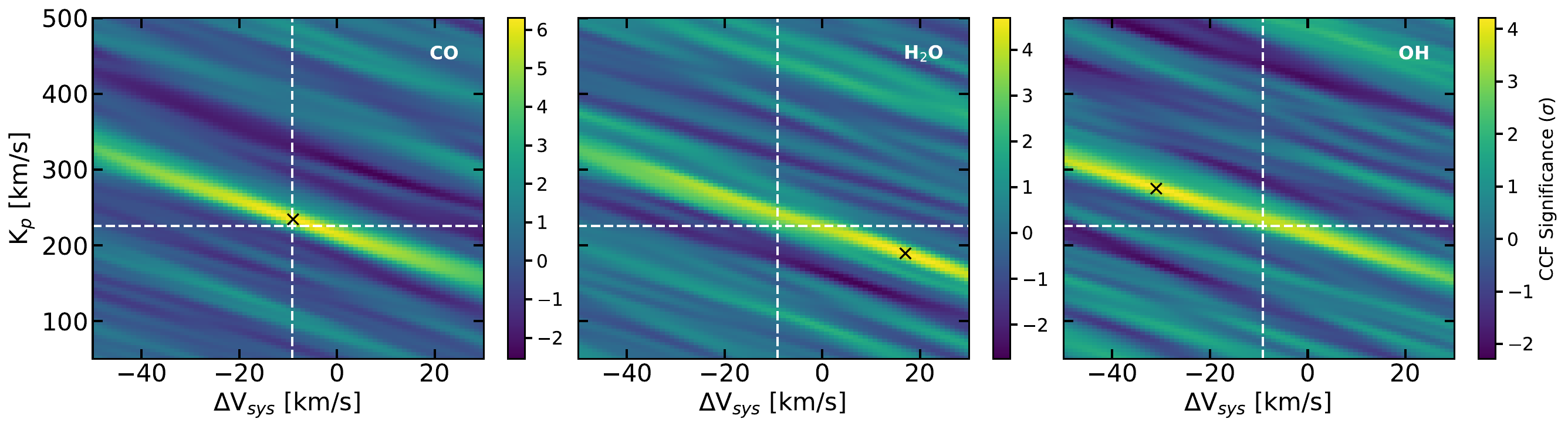}

\caption{
Cross-correlation results for CO, H$_{2}$O, and OH in WASP-33~b’s atmosphere. The detection significance reaches SNR values of 6.3 for CO, 4.7 for H$_{2}$O, and 4.2 for OH. The dashed white lines indicate the expected values for the planet’s radial velocity and the system velocity from the literature, and the black cross marks the peak value of the SNR.
}
\label{fig:CCF_indiv}
\end{figure*}

Figure~\ref{fig:CCF_all} presents the resulting cross-correlation maps, generated using the composite atmospheric model with CO, H$_{2}$O, and OH. A planetary trail before ingress is clearly shown in the top panel. The dashed red line corresponds to the Doppler shift induced by the planet’s orbital motion. The alignment of this trail with the predicted velocity curve, derived from WASP-33~b’s orbital parameters, confirms the detection of the planet’s atmospheric signal. The cross-correlation map (bottom panel in Fig.~\ref{fig:CCF_all}) shows the composite atmospheric thermal emission with SNR = 7.4. The dashed white lines indicate the expected planet's radial velocity (${\rm K}_{p}$) from F23 and a system velocity (${\rm V}_{sys}$) from \citet{Gontcharov2006}. SNR is calculated following the same approach outlined in \citet{Kasper2021} using the 3$\sigma$ clipping method. The positive cross-correlation value further suggests that the detected spectral lines are in emission rather than absorption, supporting the presence of a thermally inverted dayside atmosphere.

We also attempted to detect individual molecular signals from CO, H$_{2}$O, and OH in the atmosphere using models in Figure~\ref{fig:models}. These results are shown in Figure~\ref{fig:CCF_indiv}, formatted similarly to the result of the composite atmospheric model. CO is robustly detected at the expected planetary velocity, consistent with previous studies and confirming its presence in the planet’s dayside emission. 
With the most significant detection of CO, we tested the impact of the 71$^{st}$ order and the order selection where the CO opacity is dominant to improve the detection significance following the approach of \citet{Kanumalla2024}. Using the full order set (53 orders) applied for the other detections resulted in a CCF SNR of 5.20$\sigma$, while removing the 71$^{st}$ order yielded 5.03$\sigma$. By selecting only the orders where the CO opacity is dominant (21 orders; see Figure \ref{fig:models}), we yielded a CCF SNR of 6.30$\sigma$; without the 71$^{st}$ order, this decreased to 6.17$\sigma$. Therefore, we selected the 21-order set and validated that the impact of the 71$^{st}$ order on the detection significance is approximately 2--3\%.
The cross-correlation results of H$_{2}$O and OH exhibit velocity shifts, which may suggest additional atmospheric processes at play, as individual molecular species have shown similar shifts in previous studies \citep{Brogi2023, Cont2021, Cont2024}. One possibility is that the intense dayside heating leads to the thermal dissociation of H$_{2}$O into H and OH, resulting in non-uniform molecular distributions across the atmosphere \citep[e.g.,][Finnerty et al., in review]{Lothringer2018, Parmentier2018, Nugroho2021}. Alternatively, strong atmospheric dynamics, such as high-altitude winds or circulation patterns, may contribute to the observed Doppler shifts by redistributing molecular species across different regions of the atmosphere. Follow-up observations, particularly those covering a broader orbital phase range, will be essential to confirm these velocity shifts and further investigate the planet’s atmospheric circulation and chemistry.

Notably, we achieved high-SNR molecular detections of the planetary signal (covering only 0.05 orbital phases) with just 1.46 hours of observing time. While a direct, quantitative comparison with previous studies \citep[e.g.,][]{Nugroho2021, Finnerty2023} is not appropriate due to differences in instruments and observational setups, those studies typically required longer observation times of 2–4 hours to reach comparable detection significance. Thus, our result qualitatively demonstrates the sensitivity and efficiency of IGRINS-2 for HRCCS.

\section{Summary}
In this paper, we presented an early look at the performance of IGRINS-2 with HRCCS to study the dayside atmosphere of the ultrahot Jupiter WASP-33~b.
This assessment focused on the instrument’s detectability of molecular species, such as CO, H$_{2}$O, and OH.

We successfully detected clear cross-correlation signals from these molecular species in the dayside atmosphere of WASP-33~b with a combined SNR of 7.4. The detection was achieved by cross-correlating the post-SVD residual spectra with atmospheric models convolved to match the instrumental resolution of R=45,000. The results confirmed the presence of molecular species that were consistent with previous results from other high-resolution spectroscopy studies \citep[e.g.,][]{Nugroho2021, Cont2021, Cont2022a, Cont2022b, Yan2022, Finnerty2023}, supporting the conclusion of a thermally inverted atmosphere for WASP-33~b. The detected planetary trail, which aligns with the expected radial velocity curve, further corroborates the planetary origin of the signal and underscores the precision of IGRINS-2 in tracking Doppler-shifted atmospheric features.

The cross-correlation maps demonstrate the strength of the planetary signal and its alignment with the predicted radial velocity of the planet. This not only validates IGRINS-2’s performance for HRCCS but also highlights its potential for atmospheric characterization of exoplanets. 
The clear visibility of the planetary trail during the pre-eclipse orbital phases demonstrates the capability of IGRINS-2 to detect molecular features in exoplanetary atmospheres with high sensitivity.

Our results are in good agreement with previous studies on ultrahot Jupiters \citep[e.g.,][]{Finnerty2023, Brogi2023}, further supporting the reliability of atmospheric models for these extreme environments. In particular, our detection of CO and H$_{2}$O is consistent with theoretical predictions for high-temperature atmospheres where strong molecular excitation occurs \citep{Madhusudhan2011}. 
The clear detection of OH additionally provides indirect evidence of active thermal dissociation processes, as OH is a product of H$_{2}$O dissociation under intense stellar irradiation \citep{Brogi2023}. This further highlights the chemically complex and dynamic nature of ultra-hot Jupiter atmospheres.

In conclusion, IGRINS-2 has demonstrated its ability to perform precise HRCCS on exoplanets, providing valuable insights into the atmospheric composition of WASP-33~b. This study highlights the instrument’s effectiveness in detecting molecular signatures in the thermal emission spectra of ultrahot Jupiters and establishes IGRINS-2 as a promising tool for future exoplanet studies. Further observations across different planetary systems will be necessary to explore the broader applicability of IGRINS-2 in atmospheric characterization and to extend our understanding of the diverse chemical environments in exoplanet atmospheres.

\section*{Acknowledgments}
Y.H.C. thanks Stefan Pelletier, Woowon Byun, and Jaerim Koo for improving the manuscript. 
This research was partly supported by the Korea Astronomy and Space Science Institute under the R\&D program (Project No. 2025-1-830-05, Project No. 2025-1-860-02, Project No. 2025-1-868-02) supervised by the Korea AeroSpace Administration.
This work used the Immersion Grating Infrared Spectrograph 2 (IGRINS-2) developed and built by a collaboration between Korea Astronomy and Space Science Institute (KASI) and the International Gemini Observatory.
Based on observations obtained at the international Gemini Observatory, a program of NSF NOIRLab, which is managed by the Association of Universities for Research in Astronomy (AURA) under a cooperative agreement with the U.S. National Science Foundation on behalf of the Gemini Observatory partnership: the U.S. National Science Foundation (United States), National Research Council (Canada), Agencia Nacional de Investigaci\'{o}n y Desarrollo (Chile), Ministerio de Ciencia, Tecnolog\'{i}a e Innovaci\'{o}n (Argentina), Minist\'{e}rio da Ci\^{e}ncia, Tecnologia, Inova\c{c}\~{o}es e Comunica\c{c}\~{o}es (Brazil), and Korea Astronomy and Space Science Institute (Republic of Korea).
This work was enabled by observations made from the Gemini North telescope, located within the Mauna Kea Science Reserve and adjacent to the summit of Mauna Kea. We are grateful for the privilege of observing the Universe from a place that is unique in both its astronomical quality and its cultural significance.


\bibliography{sample631}{}
\bibliographystyle{aasjournal}



\end{document}